\documentclass{emulateapj}
\slugcomment{Accepted to The Astrophysical Journal}

\usepackage{apjfonts}
\usepackage{natbib}
\usepackage{amsmath}
\usepackage{appendix}

\usepackage{hyphenat}
\newcommand{\height}{3.4in}
\newcommand{\heights}{3.3in}
\newcommand{\heightb}{4in}

\newcommand{\wth}{$w(\theta)$}
\newcommand{\wpp}{$w_{pp}$}

\newcommand{\wppth}{$w_{pp}(\theta)$}
\newcommand{\wspthz}{$w_{sp}(\theta,z)$}
\newcommand{\wspth}{$w_{sp}(\theta)$}
\newcommand{\xir}{$\xi(r)$}
\newcommand{\xiss}{$\xi_{ss}$}
\newcommand{\xissrppi}{$\xi_{ss}(r_p,\pi)$}
\newcommand{\wprp}{$w_p(r_p)$}
\newcommand{\phip}{$\phi_p$}
\newcommand{\phipz}{$\phi_p(z)$}

\newcommand{\ross}{$r_{0,ss}$}
\newcommand{\ropp}{$r_{0,pp}$}
\newcommand{\rosp}{$r_{0,sp}$}
\newcommand{\gmss}{$\gamma_{ss}$}
\newcommand{\gmpp}{$\gamma_{pp}$}
\newcommand{\gmsp}{$\gamma_{sp}$}
\newcommand{\app}{$A_{pp}$}
\newcommand{\cpp}{$C_{pp}$}
\newcommand{\asp}{$A_{sp}$}
\newcommand{\csp}{$C_{sp}$}
\newcommand{\rospgm}{$r^{\gamma_{sp}}_{0,sp}$}

\newcommand{\Cb}{{\bf C}}

\newcommand{\beq}{\begin{equation}}
\newcommand{\eeq}{\end{equation}}

\begin{document}

\title{Improving Correlation Function Fitting with Ridge Regression: Application to Cross-Correlation Reconstruction}
\author{Daniel J. Matthews and Jeffrey A. Newman}
\affil{Department of Physics and Astronomy, University of Pittsburgh, 3941 O'Hara Street, Pittsburgh, PA 15260}
\email{djm70@pitt.edu, janewman@pitt.edu}

\begin{abstract}
Cross-correlation techniques provide a promising avenue for calibrating photometric redshifts and determining redshift distributions using spectroscopy which is systematically incomplete (e.g., current deep spectroscopic surveys fail to obtain secure redshifts for 30-50\% or more of the galaxies targeted).  In this paper we improve on the redshift distribution reconstruction methods from our previous work by incorporating full covariance information into our correlation function fits.  Correlation function measurements are strongly covariant between angular or spatial bins, and accounting for this in fitting can yield substantial reduction in errors.  However, frequently the covariance matrices used in these calculations are determined from a relatively small set (dozens rather than hundreds) of subsamples or mock catalogs, resulting in noisy covariance matrices whose inversion is ill-conditioned and numerically unstable.  We present here a method of conditioning the covariance matrix known as ridge regression which results in a more well behaved inversion than other techniques common in large-scale structure studies.  We demonstrate that ridge regression significantly improves the determination of correlation function parameters. We then apply these improved techniques to the problem of reconstructing redshift distributions.  By incorporating full covariance information, applying ridge regression, and changing the weighting of fields in obtaining average correlation functions, we obtain reductions in the mean redshift distribution reconstruction error of as much as $\sim40\%$ compared to previous methods.  In an appendix, we provide a description of POWERFIT, an IDL code for performing power-law fits to correlation functions with ridge regression conditioning that we are making publicly available.
\end{abstract}


\section{Introduction}
\label{sec:intro}
Many of the cosmological measurements to be performed with future photometric surveys will require extremely well-characterized redshift distributions \citep{2006astro.ph..9591A, 2006MNRAS.366..101H, 2006ApJ...636...21M}. A key challenge will be to obtain redshift information for galaxies from large multiwavelength samples, which include objects that are too faint for spectroscopy or fail to yield secure redshifts. A conventional approach is to use large sets of spectroscopic redshifts to establish a relation between redshift and color information to calibrate photometric redshifts \citep{1995AJ....110.2655C,2008MNRAS.390..118L, 2009ApJ...695..747B, 2009MNRAS.398.2012F}. However, at the depths probed by these surveys, existing large redshift samples have all been highly incomplete, with a strong dependence of success rate on both redshift and galaxy properties \citep{2006MNRAS.370..198C}.

\citet{2008ApJ...684...88N} described a new technique for calibrating photometric redshifts (commonly referred to as photo-$z$'s) using cross\hyp{}correlations which exploits the fact that galaxies at similar redshifts tend to cluster with each other, and in \citet{2010ApJ...721..456M} (from here on referred to as MN10) we tested this technique using realistic mock catalogs which include the impact of bias evolution and cosmic variance. We showed that for objects in a photometric redshift bin (e.g., selected using some photo-$z$-based algorithm), we can recover its true redshift distribution, \phipz, by measuring the two-point angular cross\hyp{}correlation between objects in that bin with a bright spectroscopic sample in the same region of the sky, as a function of spectroscopic $z$. The cross\hyp{}correlation signal at a given redshift will depend on both the intrinsic clustering of the samples with each other and the fraction of the photometric sample that lies at that redshift (i.e., \phip). Autocorrelation measurements give information about the intrinsic clustering of each sample and can be used to break this degeneracy.

In MN10, we assumed for convenience that correlation function measurements in different angular/radial bins were completely independent. However, analytical models as well as simulations have shown that the covariance between bins is significant \citep{1994ApJ...424..569B,2005ApJ...630....1Z,2011MNRAS.414..329C}. Incorporating all available information about this covariance should provide better constraints on the correlation function parameters used in reconstructing \phipz. In this paper we improve on the methods of MN10 by accounting for this covariance. 

However, the inversion of covariance matrices calculated from relatively small sample sizes (e.g. a modest number of mock catalogs or jackknife regions) is not well behaved: modest noise in a covariance matrix can yield large variations in its inverse. We therefore also incorporate ridge regression, a method of conditioning covariance matrices (i.e., stabilizing the calculation of their inverse) which is common in the statistics literature but novel to correlation function analyses, into our methods. We will then optimize the reconstruction of \phipz\ by varying the level of this conditioning. Throughout this paper we assume a flat $\Lambda$CDM cosmology with $\Omega_m$=0.3, $\Omega_{\Lambda}$=0.7, and Hubble parameter $H_0=100h$ km s$^{-1}$ Mpc$^{-1}$; we have assumed $h$=0.72, matching the Millennium simulations used for this work, where it is not explicitly included in formulae. 

The structure of this paper is as follows.  In \S \ref{sec:datasets} we describe the catalogs and data sets used. In \S \ref{sec:methods} we provide a summary of the reconstruction technique used in \citet{2010ApJ...721..456M}, as well as a description of the changes implemented for this paper. In \S \ref{sec:optrisk} we detail the optimization of the reconstruction achieved using a risk analysis. We summarize the results and discuss the gains from ridge regression in \S \ref{sec:conclusion}.  Finally, we include an appendix describing the public release of POWERFIT, an IDL program which performs fits for power-law correlation function parameters incorporating covariance information and conditioning via either ridge regression or singular value trimming.

\section{Data Sets}
\label{sec:datasets}
To test this method, it is necessary to construct two samples of galaxies, one with known redshift (``spectroscopic'') and the other unknown (``photometric''). We have done this using mock DEEP2 Redshift Survey light cones produced by Darren Croton. A total of 24 light cones were constructed by taking lines-of-sight through the Millennium Simulation halo catalog \citep{2006astro.ph..8019L} with the redshift of the simulation cube used increasing with distance from the observer \citep{2007MNRAS.376....2K}. The light cones were then populated with galaxies using a semi-analytic model whose parameters were chosen to reproduce local galaxy properties \citep{2006MNRAS.365...11C}. Each light cone covers the range $0.10<z<1.5$ and corresponds to a $0.5\times2.0$ degree region of sky, roughly equivalent to the area of a single DEEP2 survey field.  As a consequence, we will commonly refer to the individual light cones as ``fields'' in the remainder of this paper.

The spectroscopic sample is generated by randomly selecting 60\% of objects with observed $R$-band magnitude $R<24.1$, resembling the selection probability and depth of the DEEP2 Galaxy Redshift survey. The mean number of spectroscopic objects in a single light cone is $35,574$. The other sample, referred to hereafter as the photometric sample, is constructed by selecting objects in the mock catalog down to the faintest magnitudes available, with the probability of inclusion a Gaussian with $\langle z \rangle = 0.75$ and $\sigma_z = 0.20$. This emulates choosing a set of objects which have been placed in a single photometric redshift bin by some algorithm with Gaussian errors.  Although the selection function used is a Gaussian, the net $z$ distribution of the photometric sample will not be a pure Gaussian since the redshift distribution of the mock catalog we select from is not uniform. After applying this Gaussian selection to the mock catalog, we randomly reject half of the selected objects in order to reduce calculation time. The mean number of objects in the final photometric sample over the 24 light cones is $44,053$.

The mock catalog includes both the cosmological redshift as well as the observed redshift for each object. The observed redshift incorporates the effects of redshift-space distortions \citep{1998ASSL..231..185H}, and is the redshift value used for objects in the spectroscopic sample. When plotting the redshift distribution of the photometric sample we use the cosmological redshifts for each object (differences are small). Throughout the remainder of this paper, quantities related to the spectroscopic sample, i.e. objects with known redshifts, will be labeled with the subscript '$s$', while those related to the photometric sample will be labelled with subscript '$p$'.

\section{Methods}
\label{sec:methods}

\subsection{Correlation Functions}
\label{sec:corrfunc}

The basic quantities we will use to determine redshift distributions are the real space two-point correlation function and the angular two-point correlation function. The real space two-point correlation function $\xi(r)$ is a measure of the excess probability $dP$ (above that for a random distribution) of finding a galaxy in a volume $dV$, at a separation $r$ from another galaxy, $dP=n[1+\xi(r)]dV$ \citep{1980lssu.book.....P}, where $n$ is the mean number density of the sample. The angular two-point correlation function $w(\theta)$ is a measure of the excess probability $dP$ of finding a galaxy in a solid angle $d\Omega$, at a separation $\theta$ on the sky from another galaxy, $dP=\Sigma[1+w(\theta)]d\Omega$ \citep{1980lssu.book.....P}, where $\Sigma$ is the mean number of galaxies per steradian.

To calculate correlation functions we bin the distance between object pairs in a given catalog to get the pair counts as a function of separation (either real-space separation for \xir, or $\theta$ separation on the sky for \wth).  We similarly count the number of pairs in each bin between objects in a particular catalog and those in a randomly-distributed catalog, as well as the number of pairs between two random catalogs; we then calculate correlation functions by applying the Landy \& Szalay estimator \citep{1993ApJ...412...64L}. The random catalogs are constructed to have the same shape on the sky (and in the case of $\xi(r)$, the same redshift distribution) as the corresponding data catalog, but contain objects distributed with a uniform distribution on the sky (constant number of objects per solid angle, taking spherical geometry into account). In all cases we use a random catalog with $\sim$10 times the number of objects in the corresponding data catalog.  More details of our correlation function measurement methods are provided in MN10.  

Throughout this paper we will model $\xi(r)$ as a power law, $\xi(r)=(r/r_0)^{-\gamma}$, which is an adequate assumption (within the errors for the samples considered here) from $\sim 0.5$ to $\sim 20 h^{-1}$ comoving Mpc for both observed samples to $z \sim 1.5$ and those in the Millennium mock catalogs \citep{2006ApJ...638..668C,2008ApJ...672..153C,2010ApJ...721..456M}; higher-precision measurements would likely benefit from a halo-model analysis \citep{2007ApJ...659....1Z,2009ApJ...707..554Z}.  The angular correlation function can be related to the spatial correlation function: if $\xi(r)=(r/r_0)^{-\gamma}$, then $w(\theta)=A\theta^{1-\gamma}$, where $A \propto r^{\gamma}_0$ \citep{1980lssu.book.....P}. 

However, since the observed mean galaxy density in a field is not necessarily representative of the universal mean density, measurements of \wth\ from a particular field will in general deviate from the global mean by an additive offset known as the integral constraint. To remove this effect, we fit $w(\theta)$ using a power law minus a constant, e.g. we consider models of the form $w(\theta)=A\theta^{1-\gamma}-C$, where $C$ is the integral constraint. The specific correlation measurements that we use in the calculation of \phipz\ are the autocorrelation of the spectroscopic sample as a function of separation and redshift, $\xi_{ss}(r,z)$; the angular autocorrelation of the photometric sample, \wppth;  and the angular cross\hyp{}correlation of the two samples with each other as a function of spectroscopic redshift, \wspthz. As we will describe below, the redshift distribution of the photometric sample can be determined using the fit parameters (in particular $r_0$, $\gamma$, and $A$) of each of these three correlation functions.

\subsection{Reconstructing \phipz}
\label{sec:phipz}

Modeling $\xi(r)$ as a power law, we can determine the relationship between the angular cross\hyp{}correlation function $w_{sp}(\theta,z)$ and the redshift distribution. Following the derivation in \cite{2008ApJ...684...88N} (cf. eq. 4), the cross\hyp{}correlation between the photometric sample and spectroscopic objects at redshift $z$ can be written as
\begin{equation}
\label{eq:wsp}
w_{sp}(\theta,z) = \frac{\phi_p(z)H(\gamma_{sp})r^{\gamma_{sp}}_{0,sp}\theta^{1-\gamma_{sp}}D(z)^{1-\gamma_{sp}}}{dl/dz} ,
\end{equation}
where $H(\gamma)=\Gamma(1/2)\Gamma((\gamma-1)/2)/\Gamma(\gamma/2)$ (here $\Gamma(x)$ is the standard Gamma function) and $\phi_p(z)$ is the probability distribution function of the redshift of an object in the photometric sample. The angular size distance, $D(z)$, and the comoving distance to redshift $z$, $l(z)$, are determined from the basic cosmology. We can see that since \wspthz\ $\sim$\ \phipz\ \rospgm, there is a degeneracy in the cross\hyp{}correlation signal between the redshift distribution and the intrinsic clustering of the two samples with each other (characterized by \rosp\ and \gmsp). We can estimate these cross\hyp{}correlation parameters from the autocorrelation measurements of each sample using the assumption of linear biasing, for which $\xi_{sp}=(\xi_{ss}\xi_{pp})^{1/2}$. 

We now outline the details of the particular procedures which gave the best reconstruction of \phipz\ in MN10. First, we need to calculate the autocorrelation parameters of each sample. To determine how \xiss\ evolves with redshift we bin the spectroscopic objects in redshift and measure the two-point correlation function in each $z$-bin. We calculate \xiss\ using the as-observed redshifts from the simulation, which are affected by redshift space distortions in the line-of-sight direction \citep{1998ASSL..231..185H}. To minimize this effect it is common to calculate $\xi$ as a function of the projected separation, $r_p$, and the line-of-sight separation, $\pi$, and then integrate along the line-of-sight to obtain the projected correlation function,
\begin{eqnarray}
\label{eq:wp}
w_p(r_p)&=&2\int^\infty_0\xi[(r^2_p+\pi^2)^{1/2}]d\pi  \\
\label{eq:wpa}
        &=&r_p\left(\frac{r_0}{r_p}\right)^\gamma H(\gamma) ,
\end{eqnarray}
where $H(\gamma)$ is defined following equation \ref{eq:wsp}. By measuring \wprp\ in multiple $z$-bins and fitting with equation \ref{eq:wpa}, we can determine \ross$(z)$\ and \gmss$(z)$. We found that employing a linear fit of $r_{0,ss}$ and $\gamma_{ss}$ as a function of $z$ resulted in a better recovery of $\phi_p(z)$ than using each bin's value directly. We then calculate the angular autocorrelation of the photometric sample, \wppth, and fit to obtain the parameters \app\ and \gmpp. We use the autocorrelation parameters along with an initial guess of \ropp\ to calculate $r^{\gamma_{sp}}_{0,sp}=(r^{\gamma_{ss}}_{0,ss}r^{\gamma_{pp}}_{0,pp})^{1/2}$. We found that the best results were obtained by assuming the redshift dependence of the scale length is similar for each sample, i.e. \ropp$(z)$ $\propto$\ \ross$(z)$, with an initial guess of \ropp$(z)$ $=$\ \ross$(z)$.

For the angular cross\hyp{}correlation between the two samples, \wspthz, we bin the spectroscopic sample into small bins in redshift and, in each bin, measure the cross\hyp{}correlation between objects in that $z$-bin with all objects in the photometric sample. We can then fit to obtain the parameters \asp, \gmsp, and \csp. However, we found a significant degeneracy between these parameters when fitting. To remove this degeneracy, we fix $\gamma_{sp}=(\gamma_{ss}+\gamma_{pp})/2$ in each $z$-bin, and only fit for the amplitude and integral constraint. We choose this estimate for \gmsp\ because the clustering of the samples with each other is expected to be intermediate to the intrinsic clustering of each sample. Using the resulting values of \asp\ and \gmsp, as well as the initial guess for $r^{\gamma_{sp}}_{0,sp}$, we obtain an initial guess of the redshift distribution \phipz\ using equation \ref{eq:wsp}. Using this \phipz, along with $A_{pp}$ and $\gamma_{pp}$, we can redetermine $r_{0,pp}$ using Limber's equation \citep{1980lssu.book.....P}, which we use to redetermine $r^{\gamma_{sp}}_{0,sp}$ and thus $\phi_p(z)$. This process is repeated until convergence is reached. A more detailed description of this technique, including an error analysis for the resulting reconstructions, can be found in MN10.

We have implemented an additional step in the reconstruction of \phipz\ for this paper that was not employed by MN10. For each measurement, after fixing \gmsp\ and fitting for \asp\ and \csp\ in each $z$-bin, we performed a smooth fit to the measured values of \csp$(z)$ as a function of redshift. Using the same \gmsp\ but fixing \csp\ at the predicted values for each bin, we then fit for \asp. We obtained the best results from a Gaussian fit to \csp, although simply smoothing the measured \csp$(z)$ values with a boxcar average also resulted in significant gains in reconstruction accuracy. We initially tested these techniques for MN10, but they did not improve the reconstruction, and in some $z$-bins made the reconstruction worse. However, after incorporating covariance information into our analyses, this additional step significantly reduced errors in the reconstruction of \phipz, likely because the determination of \csp\ for each redshift bin is now more accurate.

We have also made a change in the methods used to calculate average correlation measurements from multiple light cones. In MN10 this was done by summing the pair counts over all of the fields and using the total pair counts in the Landy \& Szalay estimator. However, in the course of this paper we found that this method overestimates the mean correlation by more heavily weighting those light cones which are overdense at a particular redshift: they will both contain more pairs and, generally, exhibit stronger clustering than a randomly-selected region of the universe. For this paper, we instead determine the average correlation by calculating the correlation function in each field individually and then performing an unweighted average of those measurements. This change had little effect on the autocorrelation function of the photometric sample, \wppth, mainly because the larger volume sampled meant that the density varies less from field to field. The projected autocorrelation of the spectroscopic sample, \wprp, and the cross\hyp{}correlation measurements, \wspthz, were significantly affected by this change, however, with average decreases in the correlation strength of $\sim10-20\%$.

\subsection{Fitting Parameters Using Full Covariance Information}
\label{sec:covar}
In MN10 we fit for the various correlation function parameters (\ross, \gmss, etc.) assuming that there is no covariance between measurements in different angular/$r_p$ bins. We determined best-fit parameters by performing a $\chi^2$ minimization where the errors used were given by the standard deviation of the correlation function measurements in each of the 24 mock light-cones; i.e. the fitting assumed that the relevant covariance matrices were all diagonal. However, analytical models as well as simulations have shown that the off-diagonal elements of the covariance matrix are non-negligible \citep{1994ApJ...424..569B,2005ApJ...630....1Z,2011MNRAS.414..329C}. We have confirmed this to be the case by calculating the full covariance matrices of correlation function measurements in the 24 fields.  Therefore, in MN10 we were not exploiting the full covariance information when fitting for the correlation function parameters. By incorporating this information into our fitting process, we should expect to obtain more accurate results.

In order to calculate the parameters using the full covariance matrix we used $\chi^2$ minimization as in MN10, but in this case we calculate $\chi^2$ values taking into account the covariance:
\beq
\label{eq:chisqr}
\chi^2=\mathrm{({\bf y}-{\bf \tilde{y}})^T{\bf C}^{-1}({\bf y}-{\bf \tilde{y}})}
\eeq
where \Cb\ is the covariance matrix, {\bf y} is the observed correlation function data in each bin, and $\mathrm{\bf \tilde{y}}$ is the expected value according to a given model. As an example, for \wth\ equation \ref{eq:chisqr} becomes:
\beq
\label{eq:wchisqr}
\chi^2=\left[w(\boldsymbol{\theta})-(A{\boldsymbol{\theta}}^{1-\gamma}-C)\right]^{\mathrm{T}}{\bf C}^{-1}\left[w(\boldsymbol{\theta})-(A{\boldsymbol{\theta}}^{1-\gamma}-C)\right].
\eeq

\begin{figure}[t]
\centering
\includegraphics[totalheight=\height]{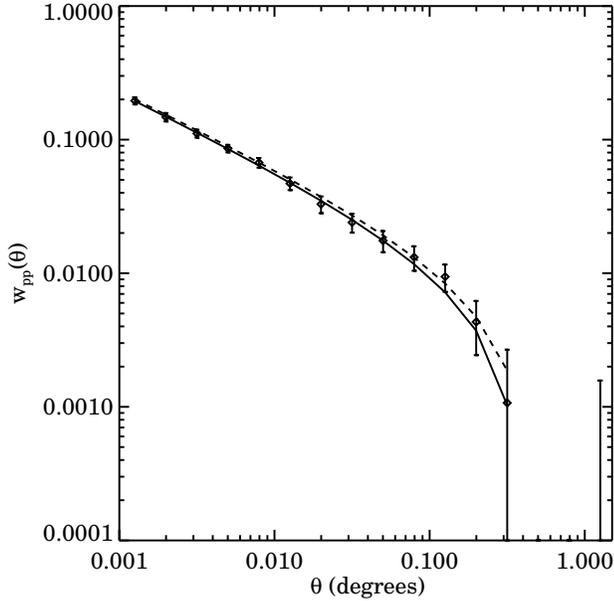}
\caption{An example of fitting a power law-integral constraint model to a measurement of the angular autocorrelation of the photometric sample, \wppth, from Millennium catalog mock light cones. The solid line is a fit assuming no covariance between angular bins, while the dashed line is a fit using the full covariance matrix, where both are fit over the range $0.001^{\circ}<\theta<1.584^{\circ}$.}
\label{fig:wfit}
\end{figure}

We start by minimizing equation \ref{eq:wchisqr} for the case of fixed $\gamma$. In that case, this minimization is simply linear regression where $\theta^{1-\gamma}$ is the independent variable, and $A$ and $-C$ are the standard 'slope' and 'intercept'. Minimizing $\chi^2$ analytically to obtain the parameters for a linear fit is  straightforward; thus for fixed $\gamma$ we can readily determine the best-fit $A$ and $C$ via standard formulae.  Alternatively, to fit for all three parameters simultaneously we can repeat the linear fit process for different values of $\gamma$, and then determine the value of $\gamma$ which minimizes $\chi^2$. We use this fitting method to determine the parameters of the angular autocorrelation of the photometric sample, \wppth, and of each $z$-bin of the angular cross\hyp{}correlation, \wspthz. For the projected real-space autocorrelation function, we see from equation \ref{eq:wpa} that $w_p(r_p)\sim r^{1-\gamma}_p$ (i.e. the same as the relation between \wth\ and $\theta$), so the fitting method is the same except that we force the intercept to be equal to zero and only fit for $\gamma$ and $A$. We then find $r_0$ using the conversion $A=r^{\gamma}_0 H(\gamma)$ from equation \ref{eq:wpa}. Figure \ref{fig:wfit} compares the fit assuming no covariance for one measurement of \wppth\ from the simulation (averaging \wpp\ from 4 of the 24 mock fields) to a fit using the full covariance matrix. 

The covariance matrices we use for fitting are calculated using correlation measurements from the 24 mock light-cones, and is therefore a sample covariance matrix and not the 'true', underlying \Cb. It can be shown that while the sample covariance matrix is an unbiased estimator of \Cb, the inverse of the sample covariance matrix is in fact a {\bf biased} estimator for the inverse of the true covariance matrix \citep{2007A&A...464..399H}. The amount of bias depends on the size of the sample used to calculate the covariance matrix; in our case, this is the number of mock catalogs (24). However, this bias can be corrected for (assuming Gaussian statistics and statistically independent measurements) simply by rescaling the inverse sample covariance matrix by a constant factor; this will not, therefore, affect the location of any $\chi^2$ minimum. We apply a bias correction where relevant in our analysis below.

\begin{figure*}[t]
\centering
\includegraphics[totalheight=\heightb]{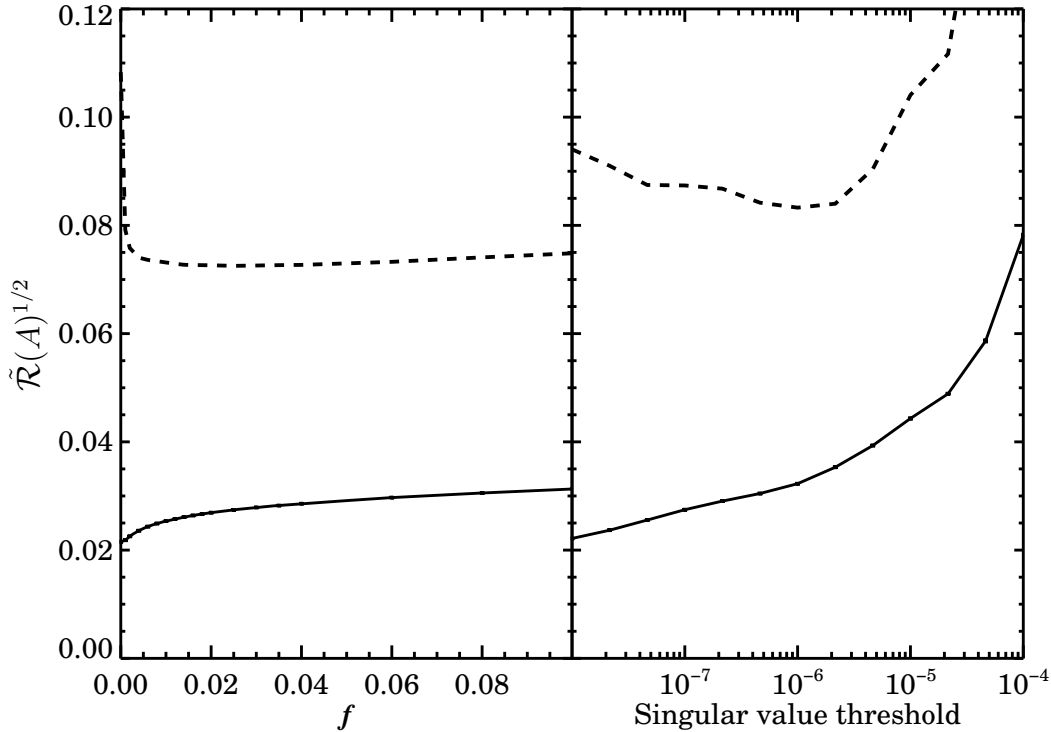}
\caption{A test of the impact of the conditioning of the covariance matrix on the results from fitting the amplitude of the correlation function, $A$.  We plot the square root of the fractional median risk (solid line) and of the maximum risk (dashed line) on $A$ as a function of the degree of conditioning.  We define the risk as the total mean squared error; i.e., the variance plus bias squared. (Left panel) We condition using ridge regression; we add a fraction $f$ of the median of the diagonal covariance matrix elements to all diagonal elements in order to stabilize the inversion of the covariance matrix.  (Right panel) We condition by inverting using singular value decomposition (SVD), setting all singular values below some threshold to zero. The median values are from a single set of $10^4$ runs, but the maximum risk line is the mean of the results from 10 sets of $10^4$ runs, as the maximum risk varied significantly from run to run.  Errors on the median are plotted, but are very small and not visible. The conditioning has a much larger effect on the maximum risk, and we therefore use a minimax optimization: i.e., choose the parameter values which make the maximum risk as small as possible. Using ridge regression, both the median and maximum optimized risk are smaller than for the SVD method. We therefore use ridge regression as our primary conditioning technique in this paper; the optimum results in fitting \wppth\ are achieved for $f\sim3\%$.}
\label{fig:arisk}
\end{figure*}

\subsubsection{Conditioning the Covariance Matrix}
\label{sec:condit}
Since we are using a covariance matrix calculated from a modest number of light cones--in effect a 'measured' covariance matrix with only a limited number of samples--noise and numerical instabilities cause difficulties when calculating $\mathrm{\Cb^{-1}}$. We found the inversion of \Cb\ to be much more well behaved when using coarser bins in $\theta$ and $r_p$ than employed in MN10. For both \wprp\ and \wth\ we doubled the bin size in log space, i.e. we use bins with $\Delta\log(r_p)=0.2$ and $\Delta\log(\theta)=0.2$. Increasing the bin size further did not yield significant improvements. 

To reduce the impact of noise in our measured covariance matrix further, we investigated several methods of conditioning the matrix (i.e., modifying the covariance matrix to improve the robustness of its inversion), and looked at how varying the conditioning improved the reconstruction. One commonly-applied method involves performing a singular value decomposition (SVD) of the covariance matrix and setting the singular values below some threshold (and their inverse) equal to zero \citep{1972GeoJI..28...97J,1972RvGSP..10..251W}. This is equivalent to performing an eigenmode analysis and trimming any unresolved modes, as is done, for instance, in \cite{2011ApJ...739...85M}. 

We also tried conditioning the covariance matrix using a technique commonly known as ridge regression \citep{1972Techno..12..69H}. This involves adding a small value to all of the diagonal elements of the covariance matrix before inverting, which reduces the impact of noise in the off-diagonal elements and makes the inversion more stable. We parameterized this conditioning by calculating the median of the diagonal elements of the covariance matrix and adding a fraction $f$ of that median value to the diagonal. We obtained better results from ridge regression than from zeroing out singular values (see \S \ref{sec:optwpp} below), and it is therefore the primary method used throughout the rest of this paper.

At first glance it may seem that applying ridge regression to the covariance matrix should be detrimental to determining the actual values of correlation function parameters: we are effectively assuming by fiat that the effective covariance matrix to be used in calculating $\chi^2$ differs from what was measured. Since ridge regression yields larger values for the diagonal elements of the covariance matrix than the data themselves would suggest, the results are equivalent to a situation with larger nominal measurement uncertainties (and hence broader $\chi^2$ minima) than implied by the original covariance matrix. 

However, when \Cb\ is determined from a limited set of measurements, $\mathrm{\Cb^{-1}}$ tends to differ significantly from the true inverse. Hence, using the standard covariance matrix in fitting should lead to measurements with nominally tighter errors than ridge regression techniques, but those measurements may in fact be significantly offset from the true value of the parameter we are attempting to determine. This can cause the parameter results to have larger spread about the true value than optimal. When we add some degree of ridge regression, the inverse of the covariance matrix is better behaved, and hence is less likely to yield a discrepant result. By varying the strength of the ridge regression conditioning, we can choose different tradeoffs between the bias and variance of parameter estimates.  In general, we want both of these contributions to be small; in the next section we investigate what degree of conditioning minimizes their sum.

\section{Risk Optimization}
\label{sec:optrisk}
In this section we will evaluate how the conditioning of the covariance matrix affects the determination of correlation function parameters and ultimately the reconstruction of \phipz.   By doing so, we will be able to optimize the reconstruction of the true redshift distribution of the photometric sample. We assess this by measuring the integrated mean squared error, i.e. the variance plus the bias squared. This is commonly referred to in statistics literature as the 'risk'. By focusing on the risk in some quantity we are optimizing for the minimum combined effect of variance and bias: either large random errors or large bias would lead to a large risk. We hence define the risk to be $\mathcal{R}\mathrm{(X)=\left<(X-X_{true})^2\right>}$, where $\mathrm{X-X_{true}}$ is the difference between the measured parameter value and its true value . At times we will also refer to the fractional risk of a parameter, which we define as $\mathcal{\tilde R}\mathrm{(X)=\langle(X-X_{true})^2\rangle/X^2_{true}}$. Since we utilize three different types of correlation measurements in the reconstruction of \phipz, we look at how changing the level of conditioning of the covariance matrix affects each one individually. 

\subsection{Optimizing Fits To \wppth}
\label{sec:optwpp}
We optimized the conditioning of the covariance matrix for the autocorrelation of the photometric sample using a Monte Carlo simulation where we use the covariance matrix of \wpp\ calculated from the 24 fields (i.e., the 24 different light cones) as our ``true'' covariance matrix, and then use it to generate realizations of correlated noise about a selected model. To do this we first find the eigenvalues and eigenvectors of the covariance matrix. We create uncorrelated Gaussian noise with variances equal to the eigenvalues, and then apply the transformation matrix constructed from the eigenvectors to this noise. This technique yields mock data with correlated noise corresponding exactly to the ``true'' covariance matrix (here, the covariance matrix of the 24 mock fields). For the true model we use $A_{true}=4.0\times10^{-4}$, $\gamma_{true}=1.58$, and $C_{true}=6.5\times10^{-3}$, which are approximately the mean parameters measured from the simulation. 

In MN10 we used the 24 mock light-cones to generate $10^4$ ``measurements'' by randomly selecting four fields at a time and finding the average \wth\ for those fields. In order to simulate this we used the method for generating correlated noise described above to create 24 realizations of single-field \wth\ measurements, and then generated $10^4$ randomly selected 'pick-4 measurements' from those 24 realizations; we will refer to each set of 24 new realizations (and its derived products) as a 'run' below. For each run we use the set of 24 realizations to calculate a measured covariance matrix, which will differ from the true covariance matrix used to generate the noise. The uncertainty in an estimate of the covariance matrix from the 24 realizations should be worse than the errors in realistic applications, making this treatment conservative. This is because the area covered by photometric surveys will in general be much larger than for the spectroscopic sample, which will result in a better constrained covariance matrix for the autocorrelation of the photometric sample; however, for the mock catalogs used here the spectroscopic and photometric areas are identical. The resulting 'measured' covariance matrix for a given run is then used to fit for the parameters of a power-law fit in each of that run's pick-4 measurements by minimizing $\chi^2$ (cf. equation \ref{eq:chisqr}).   For this and all other correlation function fits described herein we used the IDL code POWERFIT, which is being publicly released as an accompaniment to this paper.

We begin by evaluating how the reconstruction of the amplitude, $A$, changes as we vary the conditioning. The integral constraint exhibits similar behavior to the amplitude since it is proportional to the correlation strength; we are in any event not as concerned with the behavior of $C$ since it is essentially a nuisance parameter. For simplicity, we fix $\gamma$ at the true value for each run and only fit for $A$ and $C$. We calculate the risk on $A$ by performing $10^4$ runs, where for each run we:
\begin{enumerate}
\item Created 24 realizations of \wth\ as described above
\item Generated $10^4$ pick-4 measurements, randomly selecting four realizations at a time from the 24 and calculating their mean \wth
\item Fit each pick-4 measurement for $A$ and $C$ using the covariance matrix calculated from the 24 realizations created in step 1
\item Calculated the mean fractional risk on $A$ over the $10^4$ pick-4 measurements, $\mathcal{\tilde R}(A)=\langle(A-A_{true})^2\rangle/A^2_{true}$.
\end{enumerate}
We can perform the fits and calculate the fractional risk on $A$ while applying varying levels of conditioning on the covariance matrix. We parameterize the ridge regression conditioning using a variable $f$, which we define as the fraction of the  median value amongst diagonal elements of the covariance matrix which is added to the diagonal elements; i.e., we replace the $i,i$ element of the covariance matrix, $\mathrm{C}_{ii}$, by $\mathrm{C}_{ii} + f\times \mathrm{median}(\mathrm{{\bf C}}_{ii})$. For comparison, we also calculate the fractional risk on $A$ while varying the singular value threshold for the SVD conditioning described in \S \ref{sec:condit}, where all singular values below the threshold and their inverses are set to zero.

Figure \ref{fig:arisk} shows the square root of the median and maximum fractional risk amongst the $10^4$ runs as a function of both $f$ and the singular value threshold. In both cases we see that the conditioning has a much stronger effect on the maximum risk than it does on the median. We therefore perform a minimax optimization; i.e., we choose the conditioning that minimizes the maximum risk. Looking at the level of conditioning corresponding to this minimax optimization for each method, we see that the median and maximum risk are both smaller for the ridge regression conditioning. In addition, with the SVD method the maximum risk is much more sensitive to changes in the threshold around its optimized value. Small changes from the optimized threshold value in either direction can have a significant effect on the maximum risk, while the maximum risk curve for the ridge regression method is relatively flat in the vicinity of the optimized value. We therefore use ridge regression conditioning for the remainder of the calculations. By adding a few percent conditioning to our covariance matrix with the ridge regression method, we can significantly decrease the maximum risk without significantly worsening the median risk.  The optimized value for $f$ strikes a balance between the need for conditioning to stabilize inversion and the desire not to distort the relative impact of diagonal and off-diagonal covariance matrix elements, which would lead to inappropriate weighting of different data points in calculating $\chi^2$.

\begin{figure}[t]
\centering
\includegraphics[totalheight=\height]{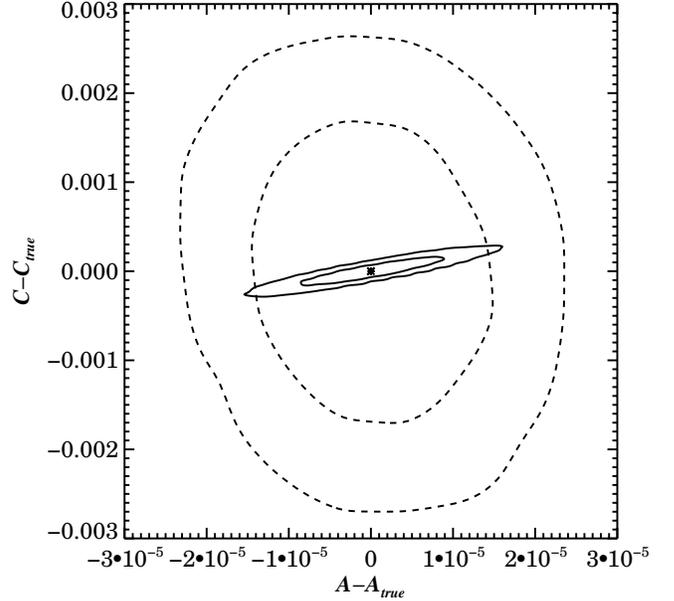}
\caption{Contour plot showing the distribution of the median values of $A-A_{true}$ and $C-C_{true}$ from each of $10^4$ runs as described in \S\ref{sec:optwpp}, where $A$ and $C$ are the fit parameters for $w(\theta)=A\theta^{1-\gamma}-C$. For our model we used $A_{true}=4.0\times10^{-4}$ and $C_{true}=6.5\times10^{-3}$. For each distribution we show the 1$\sigma$ and 2$\sigma$ contours. The solid lines are the fit parameters when using the full covariance matrix with the optimized conditioning ($f=3\%$). The dashed lines show the distribution resulting from fits with the same techniques as MN10, where we assume no covariance and fit over a smaller $\theta$ range. We are most concerned with errors in the amplitude; it is clear there is a significant improvement in the recovery of the actual value of A when the full covariance information is exploited.}
\label{fig:acplot}
\end{figure}

Figure \ref{fig:acplot} shows a contour plot of the median values of $A-A_{true}$ vs. $C-C_{true}$ amongst all pick-4 measurements for each of the $10^4$ runs using the optimized conditioning ($f=3\%$). In MN10, although we had measured the correlation function out to a separation $\theta\sim1.584^{\circ}$, we only fit over the range $0.001^{\circ}<\theta<0.1^{\circ}$. In that case, fitting over this smaller range reduced the error in $A$, and thus improved the reconstruction. When using the full covariance matrix for the fit we found that fitting over the full range of $\theta$ yielded even smaller parameter errors, as seen in Figure \ref{fig:acplot}. By utilizing covariance information in our fitting, we can robustly incorporate correlation measurements from larger scales which were useless (or even detrimental) when ignoring the covariance.

\subsection{Optimizing Fits To \wprp}
\label{sec:optwprp}
We used a different method to optimize the conditioning for the projected correlation function of the spectroscopic sample. As described in \S\ref{sec:datasets}, this sample was constructed by selecting 60\% of the objects with $R < 24.1$. We calculated the risk for the autocorrelation parameters by creating multiple samples where a different 60\% of the objects are chosen each time, and comparing these to the results for a sample containing 100\% of the objects. This differs from the method used in \S\ref{sec:optwpp} in that we are actually performing the correlation measurements using the simulations rather than generating model noise based on a covariance matrix calculated from the simulation.  In the case of \wppth\, it was more difficult to determine the true values of \wth\ (required for calculating the risk) to significantly greater accuracy than individual measurements, and therefore we relied on synthetic techniques for that analysis. Here, we have a 'truth' measurement which is much better than the fits resulting from any set of 60\% of the bright objects in only four fields, so we can measure the risk robustly without relying on simulated measurements. When calculating the reconstruction of \phipz\, we measure the parameters of a fit to \wprp\ in multiple redshift bins. For simplicity, in this section we focus on a single $z$-bin in the middle of the redshift range, $0.613 < z < 0.704$; we expect similar results for the other redshift bins.

To begin we generate $10^4$ pick-4 measurements of \wprp\ from the full sample and fit each measurement to the functional form given in equation \ref{eq:wpa}, employing the full covariance matrix calculated from the 24 fields to determine $r_0$ and $\gamma$. As in MN10, we fit over the range $0.1<r_p<10\ h^{-1}$Mpc. Since the covariance matrix calculated from the full sample should be more stable than for the 60\% subsets due to its smaller noise, we initially performed the fits with zero conditioning and used that as our 'truth'. The median values of the parameter measurements for the full sample amongst the 24 different fields were used as estimates of the true parameter values. We then calculate the risk on $r_0$ and $\gamma$ by performing 100 runs, where for each run we:
\begin{enumerate}
\item Constructed samples from each of the 24 mock fields by randomly selecting 60\% of the objects with $R<24.1$
\item Generated $10^4$ pick-4 measurements, randomly selecting four fields at a time from the 24 and calculating their mean \wprp
\item Fit each pick-4 measurement for $r_0$ and $\gamma$ using the covariance matrix calculated from the \wprp\ values measured using the 24 samples constructed in step 1\footnote{In MN10, we corrected \wprp\ for the fact that \xissrppi\ is not in actuality measured to infinite line-of-sight separation. This was not done for this test, as the correction will affect the parameters of the full sample and its subsets in a similar way, so any trends in the risk should not be affected. This saved significant calculation time. 
}
\item Calculated the mean fractional risk on both parameters, i.e. $\mathcal{\tilde R}(r_0)=\langle(r_0-r_{0,true})^2\rangle/r^2_{0,true}$ and $\mathcal{\tilde R}(\gamma)=\langle(\gamma-\gamma_{true})^2\rangle/\gamma^2_{true}$, over the $10^4$ pick-4 measurements.
\end{enumerate}
In step 3 we calculate the covariance matrix from 24 fields, which is more fields than we would actually have if we were to do cross\hyp{}correlation reconstruction with current datasets at $z\sim 1$. However, it is likely comparable to the level to which we should be able to determine the covariance matrix using current-generation deep mock catalogs, particularly since fit results will be sensitive to the relative values of covariance matrix elements, but not their absolute normalization. For each run we calculate the fractional risk on both parameters for varying levels of conditioning. 

\begin{figure}[t]
\centering
\includegraphics[totalheight=\heights]{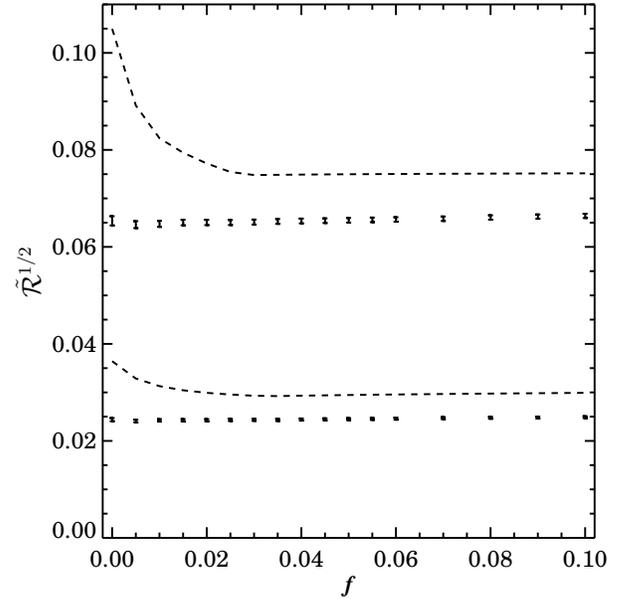}
\caption{The square root of the fractional median risk (error bars) and maximum risk (dashed line) on \ross\ (upper curves) and \gmss\ (lower curves) as a function of the degree of conditioning used for 100 runs, where 60\% of objects with $R<24.1$ were selected at random for each run, as described in \S\ref{sec:optwprp}. The conditioning has a much larger effect on the maximum risk for both parameters, and we therefore use a minimax optimization, i.e. $f$=3.5\%.}
\label{fig:rogmrisk}
\end{figure}

Figure \ref{fig:rogmrisk} shows the square root of the median and maximum fractional risk on $r_0$ and $\gamma$ amongst the 100 runs as a function of the conditioning. For both parameters we see a slight dip in the median risk over the 100 runs at $f\sim0.5\%$, but this represents only a minimal improvement. Once again we see the conditioning has a much more significant impact on the maximum risk. We optimize our fits by choosing the conditioning value that minimizes the maximum risk ($f\sim3.5\%$). 

\subsection{Optimizing \phipz\ Reconstruction}
\label{sec:optphipz}
After optimizing the fits for the autocorrelation measurements, we then looked at how conditioning the cross\hyp{}correlation covariance matrices affects the overall reconstruction of \phipz. Since the uncertainty in \phipz\ is dominated by the uncertainty in \wspthz, this conditioning should have the greatest impact on the reconstruction. We generate $10^4$ pick-4 measurements by averaging the correlation measurements from four randomly selected fields out of the 24, which we then use to calculate \phipz. For calculating the risk, we know the true redshift distribution in each field perfectly from the simulation, so we do not need to rely on synthetic techniques as in \S\ref{sec:optwpp}. Since the fits for both \wppth\ and \wprp\ were best with a few percent ridge regression conditioning (\S\ref{sec:optwpp}, \S\ref{sec:optwprp}), for simplicity we adopt $f$=3.5\% as the optimal conditioning in both cases. 

For each pick-4 measurement, we determine the autocorrelation parameters of the photometric sample by fitting the \wppth\ from the selected 4 fields using the optimally conditioned covariance matrix calculated from the 24 fields. All three parameters (\app, \gmpp, and \cpp) are left free and fit simultaneously. To measure the evolution of the correlation function parameters of the spectroscopic sample, we calculated \wprp\ in 10 $z$-bins covering the range $0.11<z<1.4$, where the size and location of each z-bin was selected such that there were approximately the same number of objects in each one. In each $z$-bin we calculate the covariance matrix from the 24 fields and fit each pick-4 measurement using the optimal conditioning to determine $r_{0,ss}(z)$ and $\gamma_{ss}(z)$.

In one redshift bin ($0.11<z<0.268$), the values of $r_{0,ss}$ and $\gamma_{ss}$ obtained with these methods were significantly different from the values determined when assuming no covariance. We investigated the likelihood contours in detail and found they were not well behaved; not only were the median parameter values different from the result with no covariance, the standard deviation of the $10^4$ pick-4 measurements proved to be an underestimate of the uncertainty in that bin, which had significant effects when performing an error-weighted linear fit to \ross$(z)$ and \gmss$(z)$. We attempted a variety of methods for estimating the errors in that bin with poor results.  However, we found that fitting over the shorter range $0.25<r_p<10\ h^{-1}$Mpc, rather than $0.1<r_p<10\ h^{-1}$Mpc, gave more well behaved values (more consistent with the values in other redshift bins or those obtained when ignoring covariance) and improved the reconstruction. For consistency we fit over this range for all bins where $z<0.8$. As in MN10 we continue to fit over the range $1.0<r_p<10\ h^{-1}$Mpc for $z>0.8$, as in the Millennium simulations (though less so in real datasets) \wprp\ diverges significantly from a power law at $0.1<r_p<1\ h^{-1}$Mpc. 

While the conditioning of the fits for the autocorrelation parameters was kept the same for each measurement, we varied the conditioning of the cross\hyp{}correlation fits to see how it affects the reconstruction. We bin the spectroscopic sample over the range $0.19<z<1.39$ with a bin size of $\Delta z=0.04$ and measure $w_{sp}(\theta)$ in each bin. At each level of conditioning we:
\begin{enumerate}
\item Calculated the covariance matrix of \wspth\ in each redshift bin from the 24 fields and apply the ridge regression conditioning to each matrix
\item Generated $10^4$ pick-4 measurements, randomly selecting four fields at a time from the 24 and calculating their mean \wspthz
\item In each $z$-bin, fit the pick-4 measurements for \asp\ and \csp, fixing \gmsp\ as described in \S\ref{sec:phipz}, using the covariance matrices calculated in step 1
\item Combined \asp$(z)$ and the optimized autocorrelation parameters for each pick-4 measurement to calculate the probability distribution function, \phipz, applying equation \ref{eq:wsp}
\item For each pick-4 measurement, we calculated the mean risk on \phipz, $\mathcal{R}(\phi_p(z))=\langle(\phi_p(z)-\phi_{p,true}(z))^2\rangle$, over the range $0.4<z<1.2$. This was done in two ways:
\begin{enumerate}
\item Using the overall mean \phipz\ of the 24 fields as $\phi_{p,true}(z)$
\item Using the mean \phipz\ from the particular 4 fields used in a given measurement as $\phi_{p,true}(z)$
\end{enumerate}
\item Calculated the mean $\mathcal{R}(\phi_p(z))$ over the $10^4$ pick-4 measurements for both types of risk

\end{enumerate}
In step 5, we calculate the risk over a slightly limited redshift range to eliminate bins where noise dominates the measurements, which diluted our ability to assess the impact of ridge regression.

\begin{figure}[t]
\centering
\includegraphics[totalheight=\heights]{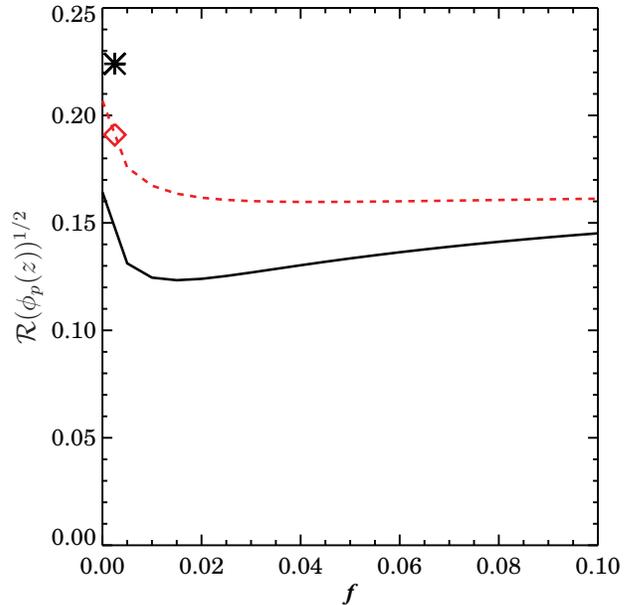}
\caption{The square root of the mean risk over the range $0.4<z<1.2$ for the reconstruction as a function of the degree of conditioning applied to the covariance matrix of \wspth\ in each redshift bin. The solid line is the risk compared to the overall mean of the 24 fields, and the star symbol is the corresponding risk using the methods of MN10. The dashed red(gray) line is the risk defined from comparing each measurement to the mean redshift distribution of the particular 4 fields used, and the red(gray) diamond symbol is the corresponding risk using the previous method. Both are at or near their minimum value at a conditioning of a few percent. The decrease in the risk when comparing to the overall mean is much greater, though improvements are significant regardless of the measure used.}
\label{fig:phirisk}
\end{figure}

\begin{figure*}[t]
\centering
\includegraphics[totalheight=6in]{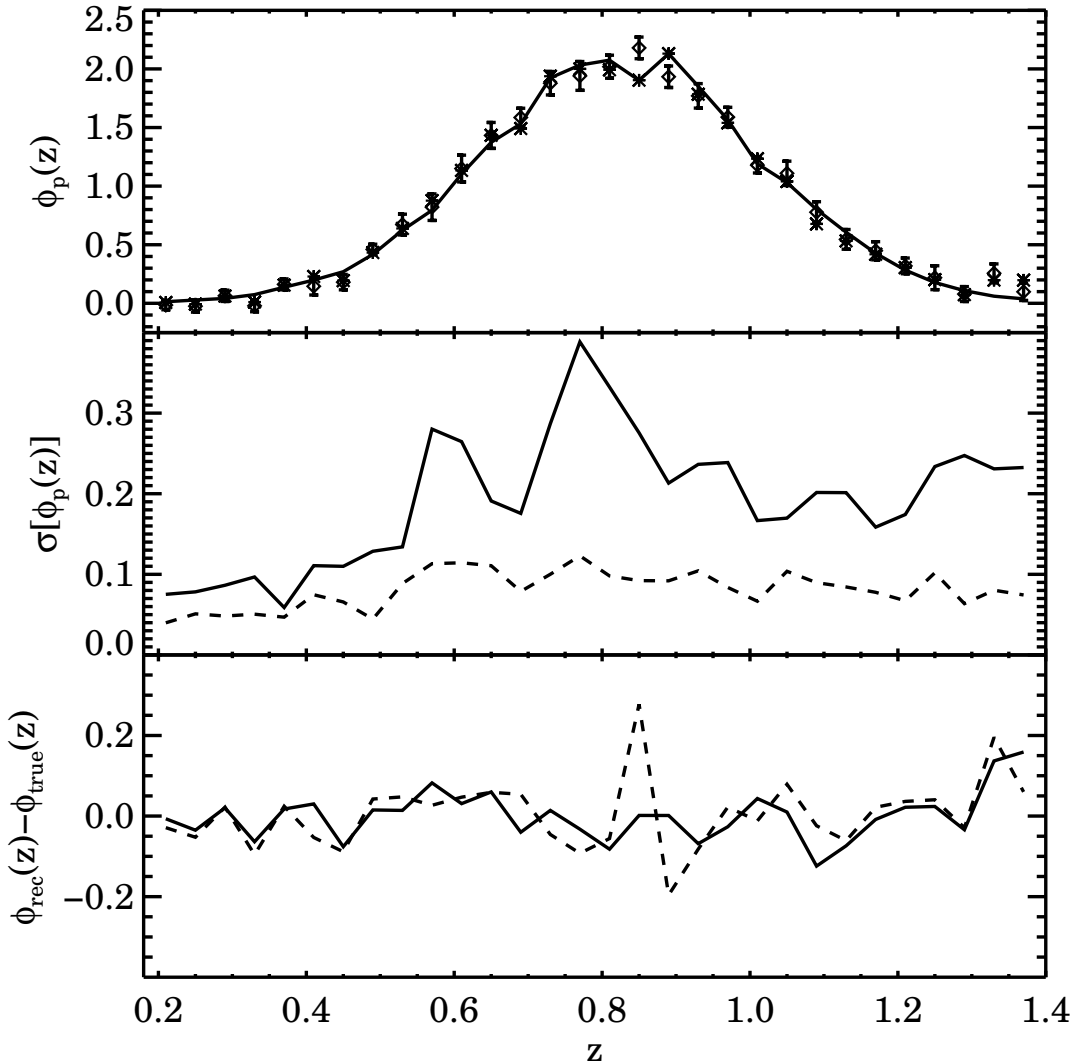}
\caption{The reconstruction of \phipz\ using 3.5\% conditioning for fits to all three correlation measurements, (i.e. \wppth, \wprp, \wspthz).  In the top panel, the solid line is the mean true distribution of the 24 fields, the star symbols are the median values of the $10^4$ pick-4 measurements obtained using the methods of MN10, and the diamonds are the median values for the optimized reconstruction using the full covariance matrix for the fits (with error bars). The middle panel compares the standard deviation of the $10^4$ pick-4 measurements in each bin using the methods from MN10 (solid line) to the improved reconstruction (dashed line), while the bottom panel compares the bias. The errors are significantly smaller in each bin, while the bias is comparable when full covariance information is used. These results are not significantly changed for moderate changes in $f$.}
\label{fig:phiplot}
\end{figure*}

Figure \ref{fig:phirisk} shows both mean risks as a function of the conditioning, compared to the risk using methods identical to MN10. We optimized for the mean risk over the redshift range rather than the maximum risk as the latter was dominated by random outliers (due to the smaller number of objects in the redshift bins used, errors in \wspthz\ are much larger, and hence random excursions extend further, than for the autocorrelations). Both techniques indicate that the minimum risk is obtained at around a few percent conditioning. There is a substantial improvement in both measures, but particularly in the risk comparing the redshift distribution for the four chosen fields to the overall (e.g. universal) mean. Figure \ref{fig:phiplot} shows the reconstruction for 3.5\% conditioning (i.e. the same for all three fits) as well as the variance and bias, and compares to the reconstruction using methods identical to MN10. The decrease in the variance is significant in each redshift bin while the bias is relatively unchanged in all but a few $z$-bins.  By incorporating full covariance information and ridge regression methods, the square root of the fractional risk is $<40\%$ smaller than that resulting from our prior methods.

\section{Summary and Conclusion}
\label{sec:conclusion}
In this paper we have improved on the cross\hyp{}correlation techniques presented in \citet{2010ApJ...721..456M} by incorporating full covariance information. In addition, we have demonstrated the improvements that result from incorporating ridge regression in fitting for correlation function parameters. Conditioning using ridge regression allowed us to obtain a more stable inversion of the covariance matrix by reducing the impact of noise in the off diagonal elements, resulting in better estimates of the correlation function parameter values; results were significantly better than with other commonly-used methods such as zeroing out small singular values in a singular value decomposition of the covariance matrix. We analyzed how this conditioning affected the integrated mean squared error, i.e. the risk, for these parameter measurements, and in doing so optimized the cross\hyp{}correlation technique for recovering the redshift distribution of a photometric sample with unknown redshifts. We also found that we gain significant improvement in the reconstruction by adding a step to the recipe described in MN10: we now perform a smooth fit for the amplitude of the integral constraint of the cross\hyp{}correlation measurements as a function of redshift, $C_{sp}(z)$.  We then refit for the amplitude of the cross\hyp{}correlation, \asp, with $C_{sp}$ fixed at the smooth fit value in each z-bin. 

We tested the effect of the ridge regression technique on the calculation of parameter values for both \wth\ and \wprp\ and found that it had a much more significant impact on the maximum risk found over multiple runs than on the median risk. In other words, it yields a great improvement in the worst-case errors, but smaller improvements in more typical cases. For \wth\ the square root of the maximum fractional risk in the amplitude, $A$, for fixed $\gamma$ decreased by $\sim35\%$ on average at a few percent conditioning. For \wprp\ we found a similar decrease for \ross\ ($\sim29\%$), while the decrease for \gmss\ was somewhat smaller ($\sim20\%$)--although still significant. After implementing the changes described above to the recipe described in MN10 we found that adding just a few percent of the ridge regression conditioning to each covariance matrix used in the calculation resulted in a significant improvement in the cross\hyp{}correlation reconstruction. When conditioning all covariance matrices at the level of 3.5\% there was $\sim42\%$ decrease in the mean of the square root of the risk on the recovered \phipz\ compared to the overall (i.e. universal) mean \phipz, and $\sim16\%$ decrease when comparing the recovered \phipz\ to the mean of the actual \phipz\ for the particular four fields used in the measurement.

In this paper, as well as in MN10, we utilized an artificial selection for objects in a photometric redshift bin that consisted simply of selecting objects with a Gaussian probability centered at a mean redshift. However, in real samples, photometric redshift errors should depend on galaxy type (and hence biasing) and will not necessarily lead to Gaussian selection probabilities or uniform evolution of bias with true redshift within the selected sample. Therefore in a future paper, we will test this improved cross\hyp{}correlation technique using these same mock catalogs but instead of assuming a redshift distribution, we will use simulated photometry to measure photometric redshifts and try to recover the redshift distribution of various photo-z selected bins. The following paper in this series will apply these techniques to test photometric redshifts for galaxies in the CANDELS multiwavelength survey \citep{2011ApJS..197...35G,2011ApJS..197...36K}.

\acknowledgements

This paper has benefited from helpful discussions with Larry Wasserman, Chris Genovese, Peter Freeman, Chad Schafer, Ann Lee, Nikhil Padmanabhan and Michael Wood-Vasey.  It was supported by the United States Department of Energy Early Career program via grant DE-SC0003960.

\appendix
\section{POWERFIT Code}
In the course of this analysis we developed a short IDL function designed to fit for the parameters of a power-law plus constant model using full covariance information, with or without conditioning of the covariance matrix. Given arrays containing the independent variable values $x$, the dependent variable values $y$, and the covariance matrix of the $y$ values, \Cb, it determines the best-fit parameters for a function of the form $y=a x^b+c$ via $\chi^2$ minimization (cf. Equation \ref{eq:chisqr}).  It outputs the best-fit parameter values in the form of a three-element array, i.e. [$a,b,c$]. POWERFIT calculates the fit parameters as described in \S\ref{sec:covar}.  If the exponent, $b$, is fixed, the best-fit values of $a$ and $c$ are calculated analytically using standard linear regression formulae. To fit for all three parameters simultaneously, POWERFIT instead uses the AMOEBA function (distributed with IDL, and based on the routine amoeba described in Numerical Recipes in C \citep{1992nrca.book.....P}) to search for the exponent value that minimizes the $\chi^2$ of the fit.

POWERFIT optionally allows the user to fix either the exponent value, $b$, the constant, $c$, or both, at specified values when calculating the fit. It is also possible to condition the covariance matrix using either of the methods described in \S\ref{sec:condit}. For ridge regression conditioning, the user must provide a value for $f$, the fraction of the median of the diagonal elements of the covariance matrix to add to the diagonal elements before inverting. For SVD conditioning, the required input is the singular value threshold; any singular values below that threshold, as well as their inverses, are set equal to zero before calculating the inversion. The code is suitable for any application where a power law or power law plus constant model is fit to data with a known covariance matrix; it can be downloaded at {\tt http://www.phyast.pitt.edu/\~{}janewman/powerfit}.

\bibliography{adsreferences}

\end{document}